\documentclass[journal]{vgtc}                     

\onlineid{1019}

\vgtccategory{Data Transformations}

\usepackage{comment}
\usepackage{multirow}
\usepackage{makecell}
\usepackage{amsmath}
\usepackage{xspace}

\usepackage{colortbl}
\usepackage{setspace}

\renewcommand{\paragraph}[1]{
\vspace{4pt}
\noindent
\textbf{#1.}}

\definecolor{appleredlight}{RGB}{255, 105, 97}
\definecolor{appleorangelight}{RGB}{255, 179, 64}
\definecolor{appleyellowlight}{RGB}{255, 212, 38}
\definecolor{applegreenlight}{RGB}{48, 219, 91}
\definecolor{applemintlight}{RGB}{102, 212, 207}
\definecolor{appleteallight}{RGB}{93, 230, 255}
\definecolor{applecyanlight}{RGB}{112, 215, 255}
\definecolor{applebluelight}{RGB}{64, 156, 255}
\definecolor{appleindigolight}{RGB}{125, 122, 255}
\definecolor{applepurplelight}{RGB}{218, 143, 255}
\definecolor{applepinklight}{RGB}{255, 100, 130}
\definecolor{applebrownlight}{RGB}{181, 148, 105}

\definecolor{applerednormal}{RGB}{255, 69, 58}
\definecolor{appleorangenormal}{RGB}{255, 159, 10}
\definecolor{appleyellownormal}{RGB}{255, 214, 10}
\definecolor{applegreennormal}{RGB}{48, 209, 88}
\definecolor{applemintnormal}{RGB}{99, 230, 226}
\definecolor{appletealnormal}{RGB}{64, 200, 224}
\definecolor{applecyannormal}{RGB}{100, 210, 255}
\definecolor{applebluenormal}{RGB}{10, 132, 255}
\definecolor{appleindigonormal}{RGB}{94, 92, 230}
\definecolor{applepurplenormal}{RGB}{191, 90, 242}
\definecolor{applepinknormal}{RGB}{255, 55, 95}
\definecolor{applebrownnormal}{RGB}{172, 142, 104}

\definecolor{applereddark}{RGB}{215, 0, 21}

\newcommand{\revise}[1]{#1}

\definecolor{applegrey}{RGB}{99, 99, 102}

\newcommand{\redd}[1]{\textcolor{applereddark}{#1}}

\definecolor{crimson}{RGB}{215,0,21}
\definecolor{lcrimson}{RGB}{255,104,97}

\def\MNC{\textsc{Mnc}\xspace}
\def\PDS{\textsc{Pds}\xspace}

\def\MNCPDS{\textsc{Pds+Mnc}\xspace}
\def\PDSMNC{\textsc{Pds+Mnc}\xspace}

\renewcommand{\paragraph}[1]{
\vspace{0.05in}
\noindent
\textbf{#1.}
}

\vgtcpapertype{algorithm/technique}

\title{
Dataset-Adaptive Dimensionality Reduction 
}

\author{%
Hyeon Jeon, Jeongin Park, Soohyun Lee, Dae Hyun Kim, Sungbok Shin, and Jinwook Seo
}

\authorfooter{
  \item Hyeon Jeon, Soohyun Lee, Jeongin Park, and Jinwook Seo (corresponding author), are with Seoul National University.
  	E-mail: \{hj, shlee\}@hcil.snu.ac.kr, \{parkjeong02, jseo\}@snu.ac.kr.
  \item Dae Hyun Kim is with Yonsei University. E-mail: dhkim16@yonsei.ac.kr.
  \item Sungbok Shin is with Aviz at Inria-Saclay and Universit\'e Paris-Saclay, Saclay, France. 
  	E-mail: sungbok.shin@inria.fr.

}

\abstract{%
Selecting the appropriate dimensionality reduction (DR) technique and determining its optimal hyperparameter settings that maximize the accuracy of the output projections typically involves extensive trial and error, often resulting in unnecessary computational overhead.
To address this challenge, we propose a \textit{dataset-adaptive} approach to DR optimization guided by \textit{structural complexity metrics}. These metrics quantify the intrinsic complexity of a dataset, predicting whether higher-dimensional spaces are necessary to represent it accurately.
Since complex datasets are often inaccurately represented in two-dimensional projections, leveraging these metrics enables us to predict the maximum achievable accuracy of DR techniques for a given dataset, eliminating redundant trials in optimizing DR. 
We introduce the design and theoretical foundations of these structural complexity metrics.
We quantitatively verify that our metrics effectively approximate the ground truth complexity of datasets and confirm their suitability for guiding dataset-adaptive DR workflow.
Finally, we empirically show that our dataset-adaptive workflow significantly enhances the efficiency of DR optimization without compromising accuracy.

}


\graphicspath{{figs/}{figures/}{pictures/}{images/}{./}} 

\usepackage{tabu}                      
\usepackage{booktabs}                  
\usepackage{lipsum}                    
\usepackage{mwe}                       
\usepackage{amssymb}

\usepackage{mathptmx}                  

\begin{document}


\firstsection{Introduction}

\maketitle

\label{sec:intro}

Dimensionality reduction (DR) is an effective tool for visualizing and analyzing high-dimensional~(HD) data \cite{mao15kdd, fu19kdd, nonato19tvcg, chatzimparmpas20tvcg, jeon25arxiv, cashman25tvcg}. 
However, DR projection of HD data inherently results in distortions leading to inaccurate representations of the original structure of the HD data (e.g., local neighborhoods or clusters) \cite{lespinats11cgf, lespinats07tnn}.
As a result, data analysis based on DR may occasionally lead to inaccurate findings or cause analysts to overlook important structural characteristics \cite{aupetit07neurocomputing, jeon25arxiv};thus, analysts should \textit{optimize} DR projections to minimize distortions of key structural characteristics \cite{jeon25chi}.

Optimizing the hyperparameters of DR techniques to minimize distortions is typically a computationally expensive process (\autoref{sec:conventional}).
Different hyperparameter settings must be tested iteratively, but the optimal number of iterations is rarely clear. 
Practitioners often run more iterations than necessary, observing minimal or no reduction in distortion after a certain point. 
This is because, unlike typical machine learning optimization such as gradient descent, DR hyperparameter optimization does not have explicit convergence criteria.
In visual analytics, a desirable DR hyperparameter optimization involves more than just minimizing a loss function---additional constraints, such as the preservation of neighborhood~\cite{venna06nn, colange19vis} and cluster structures~\cite{jeon21tvcg, martins14cg}, must also be considered. 
Representing these diverse criteria and DR techniques together within a single mathematical formulation is challenging, making the validation of convergence also challenging.
Furthermore, identifying the optimal DR technique requires comparing multiple techniques, further complicating the optimization process.

We propose a \textit{dataset-adaptive} workflow that improves the efficiency of DR optimization. 
Building upon previous findings \cite{lee11pcs, lee14cidm} that certain patterns are more prominent in HD data, our approach quantifies the prominence of these patterns to estimate the difficulty of accurately projecting the data into lower-dimensional spaces.
We introduce \textit{structural complexity metrics} to measure these patterns, and use these scores to predict the maximum accuracy achievable by DR techniques.
The metrics thus enhance the efficiency of DR optimization by (1) guiding the selection of an appropriate DR technique for a given dataset and (2) enabling early termination of optimization once near-optimal hyperparameters have been reached, avoiding unnecessary computations.

While existing metrics, such as intrinsic dimensionality metrics (\autoref{sec:intdim}), can potentially serve as structural complexity metrics, they lack the desired characteristics necessary for effective integration into our dataset-adaptive workflow (\autoref{sec:definition}). We thus introduce two novel structural complexity metrics---\textit{Pairwise Distance Shift} (\PDS) and \textit{Mutual Neighbor Consistency} (\MNC)---tailored to the dataset-adaptive workflow.
\PDS characterizes the complexity of a dataset's global structure by quantifying the shift in pairwise distances \cite{lee07springer, lee14cidm}, a well-established indicator associated with the curse of dimensionality. 
\MNC, in contrast, captures the complexity of a dataset's local structure by measuring the inconsistency between two neighborhood-based similarity functions: $k$-Nearest Neighbors ($k$NN) and Shared Nearest Neighbors (SNN) \cite{ertoz02siam}.
By jointly characterizing global and local structural complexity, these metrics effectively guide the optimization of DR techniques.
We theoretically and empirically verify that our metrics capture prominently observable patterns in higher-dimensional space, ensuring reliable guidance when identifying optimal DR techniques and hyperparameter settings.

A series of experiments with real-world datasets confirm the effectiveness of our structural complexity metrics and the dataset-adaptive workflow.
First, we verify that \PDS, \MNC, and their ensemble 
produce scores that highly correlate with ground truth structural complexity approximated by an ensemble of multiple state-of-the-art DR techniques, significantly outperforming baselines such as intrinsic dimensionality metrics (\autoref{sec:intdim}). 
Second, we verify our metrics' utility in supporting the dataset-adaptive workflow of finding optimal DR projections. 
Finally, we show that the dataset-adaptive workflow significantly reduces the computational time required for DR optimization without compromising projection accuracy. 

\section{Background and Related Work}

Our work is relevant to three areas of previous work: 
(1) DR and its distortions,
(2) dataset-adaptive approaches in machine learning, and
(3) intrinsic dimensionality metrics.

\subsection{Dimensionality Reduction and Distortions}

\label{sec:relworkva}

DR techniques abstract HD data in a 2D space while preserving important structural characteristics.
\revise{Formally, for a given HD dataset $X = \{x_i \in \mathbb{R}^D, i = 1, 2, \ldots, N\}$, DR aims to produce $Y=\{x_i \in \mathbb{R}^2, i = 1, 2, \ldots, N\}$ that minimize the structural difference between $X$ and $Y$.}
For example, while PCA \cite{pearson01tf} seeks a 2D projection that maximizes the explainability of the original variance in HD data, UMAP \cite{mcinnes2020arxiv} aims to preserve neighborhood structures, revealing local manifolds.
However, DR projections inherently suffer from distortions~\cite{jeon21tvcg, nonato19tvcg, lespinats07tnn}.
Recent works~\cite{jeon21tvcg, jeon24tvcg, stahnke16tvcg, fu19kdd, jeon25chi} show the importance of mitigating distortions in achieving reliable data analysis.

A common approach to minimize distortions is to find an effective DR technique. 
The literature has proposed various benchmark studies that compare the performance of DR techniques in producing accurate projections \cite{jeon22vis, moor20icml, fu19kdd, xia22tvcg, etemadpour15tvcg, espadoto21tvcg}. 
These benchmark studies provide guides in selecting DR techniques. 
However, they rely on limited datasets, raising concerns about generalizability.
There is no guarantee that the technique winning the benchmark will also outperform others on a new, unseen dataset.


Another strategy to reduce distortions is to search for optimal DR projections for a given dataset. 
This is done by repeatedly evaluating the accuracy of projections while testing various DR techniques and hyperparameter settings. Various DR evaluation metrics are proposed for the purpose \cite{jeon23vis}. Local metrics (e.g., Trustworthiness \& Continuity \cite{venna06nn}) evaluate how well neighborhood structure is preserved, while global metrics (e.g., KL divergence \cite{hinton02nips}) focus on the preservation of the global distances between points.
Cluster-level metrics (e.g., Label-Trustworthiness \& Continuity \cite{jeon24tvcg}) focus on cluster structure. 
However, testing diverse DR techniques and hyperparameter settings is time-consuming~\cite{sharma23iccct, liao23arxiv}, as we require several iterations to reach local optima \cite{louppe2017bayesian}. Furthermore, as we lack a definitive way to confirm that the optimum has been reached, iterations might carry on beyond their necessary point, leading to inefficiency.


\paragraph{Our contribution}
We propose a dataset-adaptive workflow that makes DR optimization efficient.
Our approach measures how difficult it is to accurately project a given dataset in the 2D space and uses this value to predict the maximum accuracy achievable by each DR technique.
This helps analysts remove ineffective techniques from the optimization process, reducing computational demand (\autoref{fig:workflow} DW1). 
This approach also makes hyperparameter optimization faster by setting the predicted maximum accuracy as a stopping criterion (\autoref{fig:workflow} DW2).

\subsection{Dataset-Adaptive Machine Learning}


\label{sec:relprev}

Making machine learning models dataset-adaptive, i.e., measuring dataset properties to guide the use of models, has been considered an effective strategy to improve their reliability and efficiency.
We discuss how clustering, natural language processing, and computer vision fields benefit from dataset-adaptive approaches. 


In the clustering field, \textit{Clusterability} metrics \cite{ackerman09icais, adolfsson19pr, mccarthy16cl, kalogeratos12nips} are proposed to quantify the extent to which clusters are clearly structured in datasets \cite{adolfsson19pr, ackerman09icais}. 
Adolfsson et al. \cite{adolfsson19pr} show that analysts can enhance the efficiency of data analysis by using clusterability to decide whether they should apply clustering techniques or not.
The degree of alignment between clusters and class labels, i.e., \textit{Cluster-Label Matching} (CLM) \cite{aupetit14beliv, jeon25tpami}, is also studied to guide the clustering benchmark. 
Jeon et al. \cite{jeon25tpami} show that using high-CLM datasets makes the benchmark of clustering techniques produce more generalizable results.

The natural language processing field proposes metrics to predict the difficulty of datasets to be learned by machine learning models, i.e., \textit{dataset difficulty}, highlighting their importance in building the models with more reliable outputs \cite{ethayarajh22icml, wang22arxiv, byrd22acl}.
The literature shows that dataset difficulty can be used to predict model accuracy and overfitting before the training. Dataset difficulty is also verified to be effective in improving dataset quality, especially by identifying mislabeled \cite{ethayarajh22icml} or ambiguously-labeled \cite{wang22arxiv} data points.

Measuring dataset difficulty is also studied in computer vision \cite{liu11mldmpr, scheidegger21vc, mayo23nips}. 
Liu et al. \cite{liu11mldmpr} develop a dataset difficulty metric that balances the model complexity and accuracy of image segmentation models.
Meanwhile, Mayo et al. \cite{mayo23nips} propose to use human viewing time as a proxy for dataset difficulty. 
These methods support scaling up the model architecture search or reducing the gap between benchmark results and real-world performance \cite{scheidegger21vc}. 

\paragraph{Our contribution} All these works validate the effectiveness of the dataset-adaptive approach.
However, we lack discussion and validation on how DR---one of the most widely studied unsupervised machine learning domains---can be aligned with dataset properties.
This research bridges this gap (1) by demonstrating the importance of measuring the property (which is structural complexity) of HD datasets before optimizing DR projections and (2) by introducing accurate and fast structural complexity metrics.

\subsection{Intrinsic Dimensionality Metrics}

\label{sec:intdim}

To the best of our knowledge, intrinsic dimensionality metrics \cite{fukunga71toc, kak20sr, bac21entropy} are currently the only available options for measuring how intricate the structure of a given dataset is. 
Projection-based intrinsic dimensionality metrics identify the dimensionality in which further increments of dimensionality may slightly enhance the accuracy of a particular DR technique. PCA is widely used for this purpose \cite{espadoto21tvcg, fukunga71toc}.
On the other hand, geometric metrics (e.g., fractal dimension \cite{karbauskaite16ifm, theiler90josaa, falconer04wiley}) evaluate how detailed the geometric structure of datasets is.
Both align well with our definition of structural complexity (\autoref{sec:definition}); intuitively, if a dataset needs more dimensionality to be accurately projected, the dataset can be considered to be more complex and tends to be less accurately projected in low-dimensional spaces.
Recent works show that intrinsic dimensionality provides a grounded basis in selecting the appropriate machine learning model to train on the datasets \cite{bac21entropy} and improving their efficiency \cite{gong19cvpr, hasanlou12grsl}.



\paragraph{Our contribution} 
Intrinsic dimensionality metrics may correlate with structural complexity by intuition.
\revise{However, }they do not satisfy the desired properties of structural complexity metrics to support the dataset-adaptive workflow.
For example, projection-based metrics suffer from low generalizability across multiple DR techniques as they depend on a specific DR technique \revise{(\autoref{sec:definition} P1)}.
\revise{Also, geometric metrics produce scores that vary with the global scaling of the dataset, compromising their applicability to real-world datasets with diverse scales \revise{(\autoref{sec:definition} P2).}}
We thus propose novel structural complexity metrics (\autoref{sec:metrics}) that satisfy these desired properties.
Our metrics outperform intrinsic dimensionality metrics in terms of accurately predicting the ground truth structural complexity (\autoref{sec:acceval}) and in properly guiding the dataset-adaptive workflow (\autoref{sec:suitability}).





\begin{figure*}
    \centering
    \includegraphics[alt={A comparative diagram illustrating the conventional and dataset-adaptive workflows for selecting and optimizing dimensionality reduction (DR) techniques such as UMAP, t-SNE, Isomap, LLE, and UMATO. On the right ("CW": Conventional Workflow), all DR techniques undergo full hyperparameter optimization for a fixed number of iterations, regardless of their effectiveness—resulting in excessive computation. On the left ("DW": Dataset-Adaptive Workflow), a predictor estimates the maximum achievable accuracy for each technique. Only the most promising technique is optimized, and the process terminates early once the predicted accuracy is surpassed—reducing iterations. A central flow shows the decision point: "Current accuracy > predicted accuracy?" A heatmap-style color bar labeled "Accuracy (the lighter, the lower)" appears at the bottom to visually support the optimization goal.},width=\linewidth]{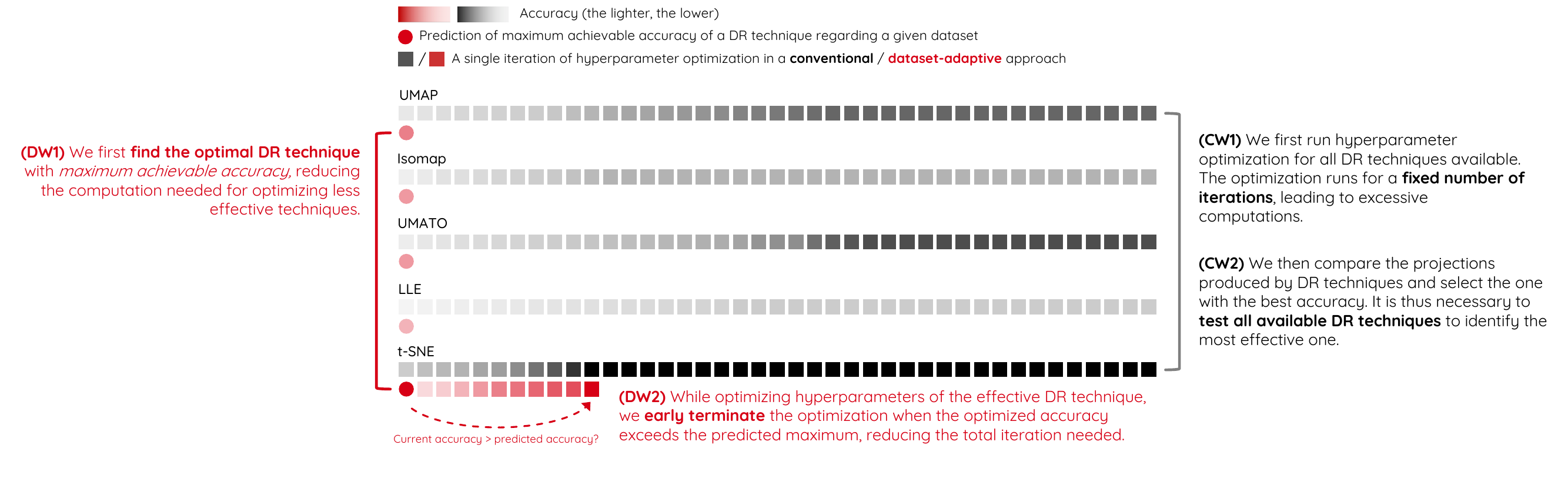}
    \vspace{-8mm}
    \caption{Illustrations of our \redd{\textit{dataset-adaptive}} workflow (DW1, DW2) and conventional workflow (CW1, CW2) of finding an optimal DR projection. 
  Each square depicts individual iterations of optimizing DR hyperparameters, where the opacity represents the maximum accuracy achieved by the current and previous iterations.  
  The dataset-adaptive workflow reduces the number of iterations required to discover the optimal projections (red squares) compared to the conventional approach (gray squares).}
    \label{fig:workflow}
\end{figure*}

\section{Conventional Workflow for Finding Optimal DR Projections}

\label{sec:conventional}

We detail the conventional workflow (\autoref{fig:workflow}CW) to find optimal DR projections with high accuracy. 
We derive the workflow by reviewing prior work that optimizes hyperparameters of DR techniques to minimize distortions \cite{jeon22vis, moor20icml, jeon24tvcg2, jiazhi21tvcg, espadoto21tvcg}. 

The conventional workflow aims to find a DR technique $T \in \mathbf{T}$ and hyperparameter setting $H \in \mathbf{H}_T$ that maximizes $C(X, T_H(X))$, where $C$ is a DR evaluation metric. $T_H(X)$ denotes the projection generated using $T$ and $H$, and $\mathbf{H}_T$ indicates the hyperparameter domain of $T$. 
\revise{Here, hyperparameter optimization often yields only marginal improvements over well-chosen default parameters \cite{espadoto21tvcg}. This means that users can naively use default hyperparameters when efficiency is crucial (e.g., when interactivity is critical). Still, optimization remains essential as it consistently ensures improved reliability in visual analytics using DR, and thus cannot be entirely avoided.}

\subsection{Workflow}

\noindent
\textbf{(Step 1) Optimizing hyperparameters of DR technique.}
In the conventional workflow, we first optimize hyperparameters $H$ for individual DR techniques $T \in \mathbf{T}$ while using $C(X, T_H(X))$ as the target function (\autoref{fig:workflow} CW1).
We do so by repeatedly testing various hyperparameter settings for a fixed number of iterations.
Here, detecting convergence in DR hyperparameter optimization is nontrivial, given the difficulty of unifying the DR technique and execution metric into one differentiable expression.
Therefore, hyperparameter search methods for non-differentiable functions, such as grid search, random search \cite{bergstra12jmlr}, or Bayesian optimization \cite{snoek12nips}, are commonly used.

\paragraph{(Step 2) Selecting the projection with the best accuracy}
We compare the accuracy of the optimal projection achieved for each DR technique. We select the one with the highest accuracy as the optimal projection (\autoref{fig:workflow} CW2).

\subsection{Problems}
This conventional workflow is time-consuming for two reasons. 
First, the workflow should test all DR techniques available (\autoref{fig:workflow} CW2).
Second, as detecting convergence in DR hyperparameter optimization is nontrivial, the workflow relies on a fixed number of iterations.
If the number of iterations is set too low, the optimization may fail to reach an optimum; if it is set too high, computation is wasted after an optimum has been reached (\autoref{fig:workflow} CW1). 
We detail how our dataset-adaptive workflow prevents these redundant computations in \autoref{sec:datasetadaptive}.






\section{Dataset-Adaptive Workflow}

\label{sec:datasetadaptive}

We introduce the dataset-adaptive workflow that reduces the redundant computations in the conventional workflow (\autoref{sec:conventional}) while still achieving an accuracy close to what is achievable in the conventional workflow. 
The central idea is to quantify how structurally complex a given dataset is, i.e., structural complexity, and then use this value to predict the maximum achievable accuracy of DR techniques in representing the original HD dataset. 
Here, we can reduce the redundant computation in the conventional workflow by (1) focusing on DR techniques with high maximum predicted accuracy (\autoref{fig:workflow} DW1), and (2) early terminating hyperparameter optimization when the predicted maximum accuracy is reached (\autoref{fig:workflow} DW2). 

In the following sections, we first define structural complexity and structural complexity metrics. 
We then detail our dataset-adaptive workflow for optimizing DR projection.

\subsection{Structural Complexity and Complexity Metrics}

\label{sec:definition}

If HD data is structurally complex, DR techniques will inevitably distort these structures, leading to less accurate low-dimensional representations..
Accordingly, we define the \textbf{\textit{structural complexity}} of a dataset $X$ as the degree to which even the best possible 2D representation having the same cardinality as $X$ falls short in accurately capturing its structure.
Formally, for $X \in \mathbb{R}^{N \times D}$ consisting of $N$ points in $D$-dimensional space, structural complexity of $X$ is defined as $S_C(X) = - \max_{Y\in \mathbb{R}^{N\times 2}}C(X,Y)$, where $C$ is a DR evaluation metric that measures \revise{a projection} $Y$'s accuracy in representing $X$'s structural characteristics. Note that we use an additive inverse to align the formula with our definition.
We set the output dimension to two since DR is commonly visualized in 2D space in visual analytics.
Note that the structural complexity of $X$ can be explained in various aspects, depending on the target structural characteristics of DR evaluation metrics (e.g., local neighborhood structure). 
This feature is crucial in leveraging structural complexity scores to guide the dataset-adaptive workflow that comprises multiple DR techniques emphasizing distinct structural characteristics.

A \textbf{\textit{structural complexity metric}} is a function \revise{$f: \mathbb{X} \rightarrow \mathbb{R}$} that is desired to have high predictive power with the ground truth structural complexity, \revise{where for any HD data $X$, $X \in \mathbb{X}$.}
Formally, given a set of datasets $\{X_1, X_2, \cdots, X_N\}$, the predictive power of $f$ with respective to $S_C$ is defined as the degree to which $\{S_C(X_1), S_C(X_2),\cdots, S_C(X_N)\}$ can be accurately predicted by $\{f(X_1), f(X_2), \cdots, f(X_N)\}$.
\revise{To properly support the dataset-adaptive workflow,} we want our metrics to be \textbf{(P1) independent of any DR technique}. This is because depending on a certain DR technique may make metrics work properly in predicting maximum achievable accuracy for some DR techniques while working poorly for others. 
Our metrics should also be \textbf{(P2) scale-invariant}, i.e., be independent of the global scale of datasets. 
\revise{Here, global scaling denotes the operation that multiplies an arbitrary positive real number $\alpha > 0$ to all values in the dataset.}
Global scaling is a characteristic unrelated to data pattern or distribution, thus also independent of the accuracy of projections in representing HD data. 
Failing to be scale-invariant thus makes the metrics hardly support the dataset-adaptive workflow, as scores can be artificially manipulated through global scaling.
\revise{Finally, we want our metrics to be \textbf{(P3) computationally beneficial}, meaning that computing the metrics should be faster than a single run of the DR technique applied in the dataset-adaptive workflow. This requirement ensures that the workflow can achieve substantial efficiency gains when optimizing DR hyperparameters.}

\revise{Here, it is worth noting that these requirements may be incomplete. Although satisfying these requirements results in effective structural complexity metrics (\autoref{sec:acceval}, \ref{sec:suitability}, and \ref{sec:improvement}), it remains uncertain whether they are the only requirements needed to characterize the metrics. Identifying new requirements that contribute to designing improved structural complexity metrics would be an interesting avenue to explore.}

\vspace{4pt}
\textit{Validity of our desired properties in the lens of existing dataset property metrics.}
The fact that \revise{P1--P3} are established as the requirements for existing dataset property metrics (\autoref{sec:relprev}) further confirms their validity \cite{adolfsson19pr, jeon22arxiv2, ackerman09icais}. Adolfsson et al. \cite{adolfsson19pr} state that P1, P2, \revise{and P3} as requirements for clusterability metrics. Jeon et al. \cite{jeon22arxiv2} state that P2 \revise{and P3} is an important requirement for CLM metrics.

\vspace{4pt}
\textit{Violations of the desired properties by \revise{widely used} intrinsic dimensionality metrics.}
Intrinsic dimensionality metrics are natural candidates for being structural complexity metrics (\autoref{sec:intdim}). However, \revise{commonly used intrinsic dimensionality metrics} do not meet our desired properties (P1 and P2) and thus have limited capability in supporting the dataset-adaptive workflow, justifying the need to design new metrics.
For example, two existing methods to compute geometric metrics (correlation method \cite{grassberger83pnp} and box-counting method \cite{liebovitch89pla}) are not scale-invariant (P2; see Appendix A for proof). In terms of projection-based metrics, they are dependent on DR techniques by design (P1), making them weakly correlate with the ground truth and less applicable to DR optimization involving multiple techniques. 
For example, if DR techniques that focus on global structures like PCA are used, projection-based metrics may inaccurately predict the maximum achievable accuracy of local techniques like $t$-SNE or UMAP. 
We empirically show that these metrics fall behind our proposed metrics (\autoref{sec:metrics}) in correlating with ground truth structural complexity (\autoref{sec:acceval}) and also show that they are not suitable for the dataset-adaptive workflow (\autoref{sec:suitability}).

\subsection{Workflow}

\label{sec:daworkflow}

We detail our dataset-adaptive workflow.

\paragraph{Pretraining}
The workflow requires a pretraining of regression models that predict the maximum achievable accuracy of DR techniques from structural complexity scores. 
We first prepare a set of HD datasets $\mathbf{X}$, a set of structural complexity metrics $\mathbf{f}$, a set of DR techniques $\mathbf{T}$, and a DR evaluation metric $C$. 
Then, for each dataset $X \in \mathbf{X}$, we compute $f(X)$ for all structural complexity metrics $f \in \mathbf{f}$, and also compute the optimal accuracy achievable by each technique $T \in \mathbf{T}$ by optimizing hyperparameters $H$ while using $C(X, T_H(X))$ as target function, where we denote this value as $\hat S_{C, T}(X)$.
Finally, for each $T \in \mathbf{T}$, we train a regression model that predicts $\hat S_{C, T}(X)$ from $\{f(X) | f \in \mathbf{f}\}$.

\paragraph{(Step 1) Finding effective DR techniques}
Given an unseen dataset $X'$, we start by predicting $\hat S_{C, T}(X')$ for each 
technique $T \in \mathbf{T}$ from $\{f(X') | f \in \mathbf{f}\}$ (\autoref{fig:workflow} DW1). We then optimize the hyperparameters 
of only the techniques that have high predicted $\hat S_{C, T}(X')$, eliminating the 
redundant computations in running hyperparameter optimization on potentially less effective DR techniques. 

\paragraph{(Step 2) Early terminating hyperparameter optimization}
We then optimize the hyperparameters for a selected method $T$, where
we halt the iteration early if the accuracy reaches $\hat S_{C, T}(X)$.
This early termination prevents unnecessary optimization iterations that would likely result in a minimal gain in improving accuracy (\autoref{fig:workflow} DW2). 
Note that we set the threshold exactly at  $\hat S_{C, T}(X)$ to minimize the risk of sacrificing accuracy; choosing a lower threshold (e.g., 95\% of  $\hat S_{C, T}(X)$) will increase the efficiency but at the cost of further accuracy loss.

\section{Structural Complexity Metrics for Dataset-Adaptive Workflow}

\label{sec:metrics}

We introduce \revise{two} structural complexity metrics---\textit{Pairwise Distance Shift} (\PDS), \textit{Mutual Neighbor Consistency} (\MNC)---that are tailored for the dataset-adaptive workflow. 
These metrics quantify phenomena that become more pronounced in higher-dimensional spaces as a proxy for structural complexity. 
\PDS estimates the complexity of global structure such as pairwise distances between data points, and \MNC focuses on local neighborhood structures.
\PDS and \MNC thus complement each other in estimating ground truth structural complexity (\autoref{sec:acceval}) and guiding dataset-adaptive workflow (\autoref{sec:suitability}), which motivates us to propose their ensemble (\PDSMNC) as an additional metric. 

\revise{Note that these two metrics depend solely on the distance matrix of the input data. This means that any HD data, regardless of its type (e.g., image, tabular), can be applied as long as the distances between pairs of points can be defined.}



\subsection{Pairwise Distance Shift (\PDS)}

\label{sec:pds}

\PDS estimates the degree of shift made by pairwise distances between data points \cite{lee07springer, lee14cidm} (also known as distance concentration \cite{francois07tkde}) as a proxy for complexity regarding global structure.
The shift refers to a common phenomenon of HD space in which pairwise distances tend to have a large average and small deviation, shifting its distribution to the positive side \cite{francois07tkde, lee11pcs}. 
Since it is naturally easier to create complex, irreducible patterns with more dimensions (Theorem 4 in Appendix A), we design \PDS to use the degree of shift as a target measurand.

\paragraph{Theoretical and empirical verification} 
Theorem 1 (Appendix A) confirms that distance shift happens more in HD spaces when the dataset is independent and identically distributed (i.i.d.); thus, measuring the shift in pairwise distances provides an effective proxy for structural complexity.
We also empirically verify the existence of distance shift in Appendix A.2.


Note that as we rely on the assumption that the data follows i.i.d. density functions, the phenomenon may not be generalized to real-world datasets. We thus conduct experiments with real-world datasets (\autoref{sec:acceval}, \ref{sec:suitability}) to verify the practical applicability of our metrics. 


\subsubsection{Algorithm}
The detailed steps to compute \PDS are as follows.

\paragraph{(Step 1) Compute pairwise distances}
For a given HD dataset $X$, we first compute the set of pairwise distances $D(X)$.

\begin{figure}
    \centering
    \includegraphics[alt={A comparison chart with two histograms side-by-side, titled “With Log” and “Without Log.” Both histograms display the distribution of data values ranging approximately from -5 to 20. The “Without Log” histogram shows a highly right-skewed distribution, where most values cluster near 0 with a long tail extending to the right. The “With Log” histogram shows a more symmetric distribution, indicating that applying a logarithmic transformation reduces the skewness of the data.}, width=\linewidth]{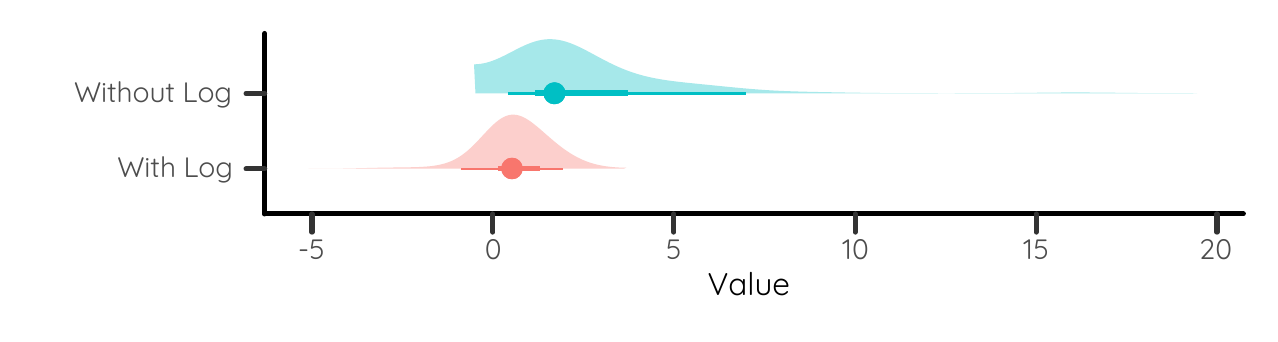}
    \vspace{-7mm}
    \caption{The distribution of \PDS scores across 96 datasets with and without log transformation. \PDS is highly skewed without log transformation.}
    \label{fig:skewness}
\end{figure}

\paragraph{(Step 2) Compute pairwise distance shift}
We then compute the final score representing the degree of shift as:
\[
\PDS = \log \mathbf{\sigma}(D(X))/\mathbf{E}(D(X)),
\]
where $\mathbf{\sigma}(D(X))$ and $\mathbf{E}(D(X))$ represent the standard deviation and the average of $D(X)$, respectively. 
Without log transformation, the distribution of \PDS scores is highly skewed (\autoref{fig:skewness}), which can harm its applicability \cite{fernandez98jasa}. 
As $0 < \mathbf{\sigma}(D(X)) < \infty$ and $0 <\mathbf{E}(D(X)) < \infty$, $-\infty < \PDS < \infty$. However, as $\mathbf{\sigma}(D(X))$ is typically less than $\mathbf{E}(D(X))$, \PDS mostly ranges from $-\infty$ to $0$. Lower scores indicate more distance shift, i.e., higher structural complexity.




\subsubsection{Compliance with the Desired Properties}

\label{sec:pdscompile}

We confirm that \PDS complies with our desired properties for structural complexity metrics (P1--P3).

\paragraph{(P1) Independence to any DR techniques}
The metric only relies on Euclidean distance, thus independent of a specific DR technique.

\paragraph{(P2) Scale-invariance}
\revise{For any scaling factor $\alpha > 0$ and dataset $X$, we have $D(\alpha X) = \alpha D(X)$. Here, we obtain:
\begin{align*}
\text{\textsc{Pds}}(\alpha X) &= \frac{\sigma(D(\alpha X))}{E(D(\alpha X))} = \frac{\sigma(\alpha D(X))}{E(\alpha D(X))} \\&=  \frac{\alpha \sigma(D(X))}{\alpha E(D(X))} = \frac{\sigma(D(X))}{E(D(X))} = \text{\textsc{Pds}}(X).
\end{align*}
Therefore, PDS is scale-invariant.}

\paragraph{\revise{(P3) Computational benefit}}
The computations of $D(X)$, $E(D(X))$, and $\sigma(D(X))$ are $O(N^2)$.
However, each step can be fully parallelized, making \PDS to \revise{highly scalable} (see \autoref{sec:imp}, \ref{sec:acceval}). \revise{We empirically verify that \PDS is computationally beneficial, i.e., faster than a single run of typical DR techniques like $t$-SNE and UMAP, in Appendix D.}

\subsection{Mutual Neighbor Consistency (\MNC)}

\MNC computes the consistency between $k$-Nearest Neighbors ($k$NN) similarity \cite{dong11www} and Shared Nearest Neighbors (SNN) similarity \cite{jarvis73toc, liu18infosci} as a proxy for the structural complexity of local structure. 
While $k$NN similarity considers a point and its $k$NN to be similar, SNN similarity regards the pair of points that share more $k$NN to be more similar. 

\MNC relies on the phenomenon that higher-dimensional spaces tend to make datasets exhibit greater inconsistency between $k$NN and SNN. 
This phenomenon originates from the pairwise distance shift. For any point $p$ in a HD dataset, the distance shift makes all other points equidistantly far from $p$, distributed on a thin hypersphere $C_p$ centered on $p$. Consequently, all other points are equally likely to be a $k$NN of $p$ with probability $k/(N-1)$, where $N$ denotes the number of points in the dataset. Now, consider $p$ and its $k$NN $q$. To make $p$ and $q$ to become SNN, a point $r$ that (1) stays within the intersection of $C_p$ and $C_q$, and (2) is a $k$NN of both $p$ and $q$ should exist. 
The first condition has a low probability of occurring as $C_p$ and $C_q$ are very thin. The second condition is also unlikely to be satisfied since the probability is $k^2/(N-1)^2$, where $N \gg k$. 
Therefore, lower $k$NN-SNN consistency likely indicates high dimensionality, which increases the likelihood of complex or irreducible patterns. We thus design \MNC to target this consistency as a key measurand.

\paragraph{Theoretical and empirical verification} 
We theoretically and empirically verify that the inconsistency between $k$NN and SNN becomes higher in HD spaces when the dataset is i.i.d. (Theorem 2 in Appendix A, Appendix A.2), implying that measuring such inconsistency provides a proxy for structural complexity. 

\subsubsection{Algorithm} We detail the procedure of computing \MNC \revise{with a hyperparameter $k$.}

\paragraph{(Step 1) Compute the $k$NN similarity matrix}
For a given dataset $X = \{p_1,p_2, \cdots, p_{N}\}$, we first compute $k$NN matrix $M^{kNN}$, \revise{which is an adjacency matrix of $X$ that regarding that two points are connected if they are in $k$NN relationship. Formally,}
\[
M^{kNN}_{i,j} = \max(0, k-r+1) \text{ if } i\neq j \text{ else } 0,
\]
where $p_j$ is $p_i$'s $r$-th NN determined by Euclidean distance.

\paragraph{(Step 2) Compute the SNN similarity matrix}
We compute SNN similarity matrix $M^{SNN}$ by setting diagonal elements as 0 and every $(i, j)$-th non-diagonal elements as:
\[
M^{SNN}_{i,j} = \sum_{(m,n) \in N_{p_i,p_j}} (k+1-m) \cdot (k+1-n),
\]
where $(m,n) \in N_{p_i,p_j}$ if $m$-th NN of $p_i$ is identical to $n$-th NN of $p_j$.

\paragraph{(Step 3) Compute the discrepancy between $k$NN and SNN matrices}
At last, the final \MNC score of $X$ is computed as:
\[
\MNC(X) = \sum_{i \in \{1, \cdots, N\}} \cos(M^{kNN}_{i, *}, M^{SNN}_{i, *}) / N,
\]
where $M_{i,*}$ denotes the $i$-th row of $M$, and $\cos$ designates cosine similarity.
We use cosine similarity as it is invariant to the scaling of similarity matrices (P2) and works robustly regardless of $N$. As cosine similarity ranges from 0 to 1, \MNC also ranges from 0 to 1 ($0 \leq \MNC \leq 1$). Lower scores indicates more inconsistency, i.e., higher structural complexity.

\subsubsection{Compliance with the Desired Properties}

\label{sec:mnccompile}

We discuss how \MNC meets our desired properties for structural complexity metrics (\autoref{sec:definition}).



\paragraph{(P1) Independence to any DR techniques}
\MNC relies only on distance functions ($k$NN and SNN) and is independent of any DR technique. 

\paragraph{(P2) Scale-invariance}
\MNC scores do not change due to the global scaling of datasets as both $k$NN and SNN are scale-invariant. 
$k$NN is scale-invariant as the ranking of NNs is not affected by scaling.
 SNN is also scale-invariant as it only relies on $k$NN similarity. 

\paragraph{\revise{(P3) Computational benefit}}
$k$NN and SNN matrix construction requires $O(kN\log N)$ and $O(kN^2)$, respectively. 
However, $k$NN can be accelerated up to $O(\log^2|k|)$ \cite{johnson21tbd}, and for the SNN matrix, we can compute every matrix cell in parallel. 
Computing cosine similarity over rows can also be \revise{fully} parallelized as every pair of rows can be treated individually. \revise{Such parallelization} makes \revise{the run of \MNC to be significantly faster than typical DR techniques (empirical validation in Appendix D)}. We empirically demonstrate the practical efficiency of \MNC in our quantitative analysis (\autoref{sec:acceval}).

\subsection{\PDSMNC}

We propose to leverage the ensemble of \PDS and \MNC, which we call \PDSMNC. This can be one by using both \PDS and \MNC scores as independent variables of regression models in predicting structural complexity or the maximum achievable accuracy of DR techniques. 
As \PDS and \MNC focus on global and local structure, respectively, we can expect \PDSMNC to generally work well in practice.
We recommend to use multiple \MNC with different $k$ values as regression models theoretically work better when with more variables.



\subsection{Implementation}

\label{sec:imp}

We develop our metrics using Python. 
We use \texttt{CUDA} provided by \texttt{numba} \cite{lam15llvm} to optimize and parallelize the algorithms. We also exploit \texttt{faiss} \cite{johnson21tbd} to parallelize $k$NN computation. 
\revise{One limitation here is that we store distance matrices within the memory of a single GPU; thus, we cannot deal with large datasets (see Appendix D for details). Improving the implementation to handle larger data will be a critical future work to increase the practical applicability of the metrics.}

\section{Experiment 1: \revise{Validity of \PDSMNC}}

\label{sec:acceval}

We evaluate how accurate our structural complexity metrics can predict the ground truth structural complexity. 

\subsection{Objectives and Study Design}

We want to verify that \PDS, \MNC, and \PDSMNC have high predictive power towards ground truth structural complexity, comparing them against baselines (intrinsic dimensionality metrics).
\revise{We also aim to provide practical guidelines for selecting structural complexity metrics for different DR evaluation metrics $C$.}
We first approximate the ground truth structural complexity of HD datasets following its definition (\autoref{sec:definition}).
We then evaluate the accuracy of our structural complexity metrics and baselines in predicting the approximated ground truth. 


\paragraph{Approximating ground truth structural complexity}
The procedure of approximating ground truth structural complexity aligns with its definition (\autoref{sec:definition}).
For a given evaluation metric $C$, we approximate ground truth $S_C(X)$ by identifying a projection $Y$ that has maximal accuracy $C(X, Y)$, and using this accuracy as the ground truth structural complexity. 
Exhaustively testing every possible 2D projection guarantees finding the global optimum but requires testing an infinite number of projections. Instead, we test projections with a high probability of being local optima to approximate $Y$.
We do so by leveraging the ensemble of multiple DR techniques. We prepare DR techniques and identify the optimal accuracy of each technique using Bayesian optimization \cite{snoek12nips}.
Then, the maximum optimal accuracy obtained by DR techniques is set as an approximated ground truth structural complexity.

For DR techniques, we first use $t$-SNE, UMAP, LLE \cite{roweis00science}, and Isomap \cite{tenenbaum00aaas}, which are currently (Mar 2025) the most widely referenced DR techniques according to their citation numbers in Google Scholar. We also use PCA as a representative global DR technique. 
Finally, we use UMATO \cite{jeon22vis} to add more diversity, as the technique aims to balance the preservation of local and global structures. Refer to Appendix B for the technical details, e.g., hyperparameter settings. 

For $C$, we use five representative DR evaluation metrics \cite{jeon23vis} that quantify the distortions focusing on diverse structural characteristics.
We pick two local metrics (T\&C, MRRE), two global metrics (Spearman's $\rho$ and Pearson's $r$), and one cluster-level metric (Label-T\&C). 
For T\&C, MRREs, and Label-T\&C, we use F1 score of two scores produced by the metric. 
Note that we constrain $C$ to satisfy P2 to make their scores comparable across datasets. For example, Steadiness \& Cohesiveness \cite{jeon21tvcg}, KL divergence \cite{hinton02nips},  and DTM \cite{chazal11fcm} are excluded as they are not independent of global scaling (proofs in Appendix A). Please also refer to Appendix B for the technical details.

\paragraph{HD datasets}
We use 96 real-world HD datasets \cite{jeon25tpami} having diverse characteristics (e.g., number of points, dimensionality). For the datasets exceeding  $3,000$ points, we randomly select a sample of 3,000 points. This is because computing approximated ground truth requires too much time (e.g., more than a week) for some datasets.


\paragraph{Baselines}
We use both projection-based and geometric intrinsic dimensionality metrics (\autoref{sec:intdim}) as baselines. We use the PCA-based method for the projection-based metric due to its efficiency (P3) and popularity in literature \cite{espadoto21tvcg, fan2010arxiv, gong19cvpr}. We compute the number of principal components required to explain more than 95\% of data variance, following Espadoto et al. \cite{espadoto21tvcg}. 
For the geometric metric, we use the correlation method due to its robustness in HD compared to the box-counting method \cite{grassberger83pnp} (Appendix C).

\paragraph{Measurement}
We evaluate how well our metrics and baselines correlate with the approximated ground truth structural complexity by computing the capacity of regression models to predict the approximated ground truths from metric scores.
This is done by quantifying average  $R^2$ correlation scores obtained by five-fold cross-validation.
As we only have 96 datasets, the results may depend on how datasets are split for cross-validation. 
Therefore, for each metric, we repeat the measurement 100 times with different splits and report the maximum score, i.e., the ideal correlation.  

To avoid bias, we use five regression models: linear regression (LR), Polynomial regression (PR), $k$NN regression ($k$NN), Random forest regression (RF), and Gradient boosting regression (GB). Please refer to Appendix B for technical details.

\paragraph{Hyperparameters}
We set $k=50$ for \MNC; we show the robustness of \MNC against varying $k$ in Appendix E.
In terms of \PDSMNC, theoretically, adding more variables always increases the predictive power of regression models. 
However, as we have a small number of datasets, adding more variables can make the models suffer from data sparsity. Our ensemble thus consists of \PDS and three \MNC{}s ($k=25, 50$, and 75).
All other competitors do not have hyperparameters.

\paragraph{Apparatus}
We execute the experiment using a Linux server with 40-core Intel Xeon Silver 4210 CPUs, TITAN RTX, and 224GB RAM. 
We use a single GPU to compute complexity metrics. 
We use this machine also for the following experiments.

\definecolor{red}{RGB}{250, 95, 126}
\definecolor{lightred}{RGB}{247, 163, 180}
\definecolor{lightlightred}{RGB}{245, 208, 216}

\definecolor{myblue}{RGB}{66, 135, 245}
\definecolor{mylightblue}{RGB}{121, 168, 242}
\definecolor{mylightlightblue}{RGB}{197, 213, 240}

\newcommand{\redc}{\cellcolor{red}}
\newcommand{\lredc}{\cellcolor{lightred}}
\newcommand{\llredc}{\cellcolor{lightlightred}}

\begin{table}[t]
    \centering
    \caption{Results of our analysis on the accuracy of structural complexity metrics in predicting the ground truth (\autoref{sec:acceval}). Each cell depicts the correlation between structural complexity metrics or intrinsic dimensionality metrics (columns) and ground truth structural complexity approximated using five DR evaluation metrics (rows).
    \PDS and \MNC show high correlations with global and local structural complexity, respectively, outperforming intrinsic dimensionality metrics. Their ensemble (\MNCPDS) achieves the best correlation for all cases. 
    }
    \scalebox{0.92}{
    \begin{tabular}{crccccc}
    \toprule
     & & \multicolumn{2}{c}{Local} & Cluster & \multicolumn{2}{c}{Global} \\
     \cmidrule(lr){3-4} \cmidrule(lr){5-5} \cmidrule(lr){6-7} 
     & &  T\&C & MRREs & L-T\&C & S-$\rho$ & P-$r$ \\ 
     \midrule
         \multirow{5}{*}{\makecell{Int. Dim. \\ (Projection)}} & LR  & .5574 \llredc& .5133\llredc& .3364& .3629& .3576 \\
                           & PR  & .5334\llredc& .4505\llredc & .2369 & .4231 \llredc& .3769 \\
                           & $k$NN & .6353\lredc& .5537\llredc& .3770 & .6495 \lredc&.6281 \lredc\\
                           & RF    & .5815\llredc&.4710 \llredc&.3013 &.6309 \lredc& .5996 \llredc\\
                           & GB   & .6132\lredc&.4021 \llredc&.3183 &.6278 \lredc&.5652 \llredc \\
    \midrule
    \multirow{5}{*}{\makecell{Int. Dim. \\ (Geometric)}} & LR  & $<0$& .0142& .0065& .0689& .0478 \\
                           & PR  & .0123& 0.2351& .0045& .0716 &.0566 \\
                           & $k$NN & .5214 \llredc& .3656& .3755& .5826\llredc& .5967\llredc\\
                           & RF    & .5054\llredc& .3807& .3958& .6010 \lredc& .5881\llredc\\
                           & GB   & .5587\llredc& .4430 \llredc&.3822 & .5711\llredc& .5730 \llredc\\
    \midrule
    \midrule
     \multirow{5}{*}{\PDS} & LR  &  .2781& .0673& .3404& .5152\llredc& .5332\llredc\\
                           & PR  & .3440& .0866& .4032\llredc& .7304\lredc& .7164\lredc\\
                           & $k$NN & .4133\llredc& .3331& .4998\llredc& .8003\redc& .8180\redc\\
                           & RF    & .4119\llredc& .3509 & .5010\llredc & .8282\redc& .8217\redc\\
                           & GB   & .4523\llredc& .3182 & .4681\llredc& .7959 \lredc&  .8075\redc\\
    \midrule
    \multirow{5}{*}{\MNC} & LR  & .8454 \redc& .6784 \lredc& .3692& .5677 \llredc& .5241\llredc \\
                           & PR  & .8807 \redc& .7244 \lredc& .3174 & .5525 \llredc& .5140\llredc \\
                           & $k$NN & .8780\redc& .7007 \lredc& .4020\llredc & .5962\llredc& .5814\llredc\\
                           & RF    & .8706\redc& .7302\lredc& .4189\llredc& .6166\lredc& .5734 \llredc\\
                           & GB   & .8666\redc& .7207\lredc& .2827& .5992\llredc&  .5741\llredc\\
    
    \midrule
    \multirow{5}{*}{\MNCPDS} & LR  & .8513\redc& .7484\lredc& .5104\llredc& .6772\lredc& .7474\lredc\\
                           & PR  & \textbf{\textit{.8984}}\redc& \textbf{\textit{.8423}}\redc& .0015& .5954\llredc& .6572\lredc\\
                           & $k$NN & .7472\lredc& .6109\lredc& .4971\llredc& \textbf{\textit{.8290}}\redc& \textbf{\textit{.8401}}\redc\\
                           & RF    & .8881\redc& .7506\lredc& .5823\llredc& .8273\redc& .8092\redc\\
                           & GB   & .8694\redc& .7636\lredc& \textbf{\textit{.6067}}\lredc& .8280\redc&  .8079\redc\\

     \bottomrule
\addlinespace[0.115cm]
\multicolumn{7}{l}{
  1. \textcolor{red}{$\blacksquare$} / \textcolor{lightred}{$\blacksquare$} / \textcolor{lightlightred}{$\blacksquare$}: very strong ($R^2 \geq 0.8)$ / strong ($0.6 \leq R^2 < 0.8)$ / 
} \\
\addlinespace[0.05cm]
\multicolumn{7}{l}{
 \hspace{18mm}moderate ($0.4 \leq R^2 < 0.6)$ correlation \cite{sarjana20jtm}
} \\ 
\addlinespace[0.115cm]
 \multicolumn{7}{l}{
 2. \textbf{\textit{Bold and italic}} refers to the top score of the column
}
    \end{tabular}
    }
    
    \label{tab:acceval}
\end{table}

\subsection{Results and Discussions}

The followings are the findings of our experiment.

\paragraph{Validity of the design of \PDS and \MNC}
The results (\autoref{tab:acceval}) verify that \PDS and \MNC are appropriately designed and work as intended. While \PDS shows at most a very strong or strong correlation with global structural complexity (S-$\rho$ and P-$r$), \MNC achieves at most a very strong correlation with local structural complexity (T\&C and MRREs). 
Meanwhile, both metrics show relatively weak correlations in the opposite cases; showing at most strong or moderate correlations. 
\PDS also shows at most strong correlation with linear and polynomial regression models (LR, PR).
As these models do not work well with non-Gaussian data, such results verify the validity of log transformation (Step 2). 
In contrast, intrinsic dimensionality metrics correlate weakly with the ground truth. For all types of ground truth structural complexity, the best correlation achieved by these metrics is substantially lower than that achieved by $\PDS$, $\MNC$, and $\MNCPDS$. 

\paragraph{Superior predictive power of \PDS and \MNC}
The results moreover confirm the effectiveness of \PDSMNC.
\PDSMNC achieves the best correlation regardless of DR evaluation metrics used to approximate ground truth structural complexity, making substantial performance gains compared to using \MNC and \PDS individually. For example, \MNCPDS has at most a strong correlation even for cluster-level structural complexity. \MNCPDS is also the sole subject that achieves a very strong correlation for the structural complexity approximated by MRREs. 
Such results underscore the importance of testing alternative strategies to ensemble complexity metrics or identifying the most effective combination. Aligned with this finding, we use \PDSMNC to demonstrate the applicability of our metrics in supporting our dataset-adaptive workflow (\autoref{sec:suitability}).

\paragraph{Weaknesses of our structural complexity metrics}
We also reveal the weaknesses of our complexity metrics. 
\PDS and \MNC show at most a weak correlation with cluster-level structural complexity (L-T\&C). Although \MNCPDS shows at most a strong correlation, the best $R^2$ value obtained is substantially lower compared to the ones from global and local structural complexity. 
Also, \MNC has a relatively weak correlation with local structural complexity approximated by MRREs compared to the one approximated by T\&C. 
\revise{We recommend not using the structural complexity metrics for such cases. For example, we recommend using \PDS and \MNCPDS when S-$\rho$ or P-$r$ is used, but not \MNC. This indicates that we currently lack structural complexity metrics that can be reliably used for cluster-level evaluation metrics (L-T\&C). Such} results underscore the need to design new, advanced structural complexity metrics that can complement \PDS, \MNC, and \PDSMNC.

\begin{figure}
    \centering
    \includegraphics[alt={A log-scaled bar chart comparing the execution time (in seconds) of different DR evaluation methods. The x-axis shows time on a logarithmic scale from 1e-2 to 1e+4 seconds. The y-axis lists several methods: DR Ensemble, Intrinsic Dim (geo), Intrinsic Dim (proj), PDS+MNC, MNC, and PDS. The chart highlights that PDS+MNC and related methods are faster than more exhaustive ensemble or geometric-based dimensionality evaluations.},width=\linewidth]{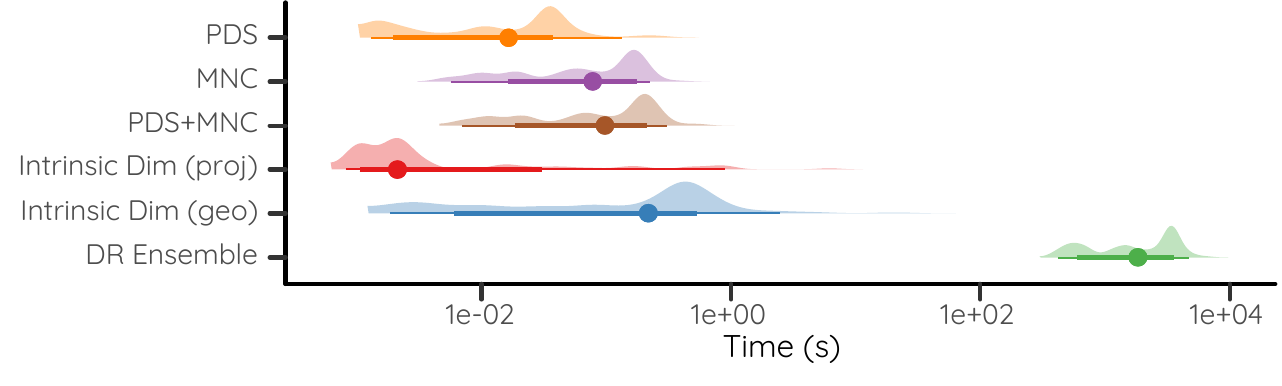}
    \vspace{-5mm}
    \caption{The runtime of our structural complexity metrics, intrinsic dimensionality metrics, and DR ensemble to produce ground truth structural complexity. 
    \PDS, \MNC, and \PDSMNC are faster than the geometric intrinsic dimensionality metric and the ensemble. The thick and thin lines indicate the 66\% and 99\% interval of probability mass.
    }
    \label{fig:efficiency}
\end{figure}

\paragraph{Computational efficiency of the metrics}
We also investigate the execution time of \PDS, \MNC, \PDSMNC, and baselines in being applied to 96 real-world datasets we use. We also examine the execution time for approximating ground truth structural complexity using the ensemble of DR techniques.
Note that we rely on GPU-based parallelization and highly optimized libraries like \texttt{scikit-learn} \cite{pedregosa11jmlr} and PyTorch \cite{paszke19neurips} while implementing intrinsic dimensionality metrics to ensure fairness.

As a result (\autoref{fig:efficiency}), we find that our structural complexity metrics are slower than the projection-based intrinsic dimensionality metric but are faster than other baselines. All three metrics required less than one second for all datasets, substantially scaling up the DR ensemble method. 
Regarding the high accuracy of \PDS and \MNC in predicting the ground truth structural complexity produced by the DR ensemble (\autoref{sec:acceval}), the result indicates that we achieve a favorable balance between efficiency and accuracy.
We further examine the efficiency of our metrics using larger datasets in Appendix D. 



\section{Experiment 2: \revise{Suitability of \PDSMNC for the Dataset-Adaptive Workflow}}

\label{sec:suitability}

We want to examine the utility of \PDSMNC, the most advanced structural complexity metrics, in properly guiding the dataset-adaptive workflow (\autoref{sec:daworkflow}).
\revise{As with a previous experiment (\autoref{sec:acceval}), the experiment also aims to provide guidelines to select structural complexity metrics in practice.}
We first evaluate the accuracy of the pretrained regression models in predicting the maximum achievable accuracy of DR techniques from the scores from \PDSMNC (\autoref{sec:maxacc}).
We then evaluate the effectiveness of \PDSMNC in guiding each step of our workflow (\autoref{sec:evalguiding} for Step 1 and \autoref{sec:evalguiding22} for Step 2).

\begin{table}[t]
    \centering
    
    \caption{Accuracy of baseline metrics and \PDSMNC on predicting maximum accuracy achievable by different DR techniques (\autoref{sec:maxacc}). \PDSMNC shows strong predictability for the majority of cases, outperforming baseline in the majority of cases. }
    \scalebox{0.91}{
    \begin{tabular}{crccccc}
         \toprule 
                &  &  T\&C & MRREs & L-T\&C & S-$\rho$ & P-$r$ \\
\midrule
               \multirow{6}{*}{\makecell{Int. Dim. \\ (Projection)}}  & UMAP    & .6540   \lredc & .6449 \lredc  &  .2656  &  .6389 \lredc &  .3690\\
         &$t$-SNE  & .7615 \lredc & .6643 \lredc & .0212 &  .5746  \llredc  &  .1674\\
        & PCA      & .8027 \redc &  .8430 \redc& .6524 \lredc  &  .7112 \lredc & .6445\lredc  \\
        & Isomap  & .7998 \lredc &    .7924 \lredc &  .6094 \lredc &  .7326   \lredc  &  .6557 \lredc \\
        & LLE     & .7747 \lredc &  .7846  \lredc & .2987 &   .6737  \lredc &  .7473 \lredc \\
        & UMATO   & .7207 \lredc & .6906 \lredc & .3222 & .7109 \lredc &  .2838 \\
\midrule
               \multirow{6}{*}{\makecell{Int. Dim. \\ (Geometric)}}  & UMAP    & .7146 \lredc  & .6519  \lredc&  .4204  \llredc &  .6550 \lredc &  .0405\\
         &$t$-SNE  & .5097 \llredc & .5767 \llredc & .0029 &  .3303  &  .1257\\
        & PCA      & .8588 \redc &  .8417 \redc & .1720 &  .7989 \lredc& .7160 \lredc\\
        & Isomap  & .8253 \redc &    .7069 \lredc&  .5629 \llredc  &  .8544   \redc &  .8855 \redc \\
        & LLE     & .8496 \redc &  .4476  \llredc & .7058 \lredc&   .7123 \lredc&  .7539 \lredc\\
        & UMATO   & .8438 \redc & .7038 \lredc& .2275 & .5660 \llredc &  .1383 \\ 
        \midrule 
        \midrule
\multirow{6}{*}{\MNCPDS} & UMAP    & .8955   \redc   & .8183  \redc&  .3944  &  .7978 \lredc &  .7578 \lredc\\
         &$t$-SNE  & .9471 \redc& .7848 \lredc& .4303 \llredc &  .8600 \redc   &  .6569 \lredc\\
        & PCA      & .9150 \redc &  .8446 \redc & .5784 \llredc &  .9063 \redc& .8836  \redc\\
        & Isomap  & .8514 \redc&    .9123 \redc &  .7962 \lredc   &  .9454 \redc       &  .8935 \redc\\
        & LLE     & .8781   \redc  &  .8501 \redc & .6986 \lredc  &   .7645 \lredc  &  .7739 \lredc\\
        & UMATO   & .9223 \redc& .7823 \lredc & .5733 \llredc & .8392 \redc &  .6880 \lredc\\
        \bottomrule
        \addlinespace[0.115cm]
         \multicolumn{6}{l}{
         \makecell{
        \textcolor{red}{$\blacksquare$} / \textcolor{lightred}{$\blacksquare$} / \textcolor{lightlightred}{$\blacksquare$}: very strong ($R^2 \geq 0.8)$ / strong ($0.6 \leq R^2 < 0.8)$ / \\ 
        \hspace{13mm}moderate ($0.4 \leq R^2 < 0.6)$ predictive power\cite{sarjana20jtm}
        }
        }
    \end{tabular}
    }
    \label{tab:indmetric}
\end{table}

\subsection{Evaluation on Pretraining Regression Models}

We detail our evaluation on the suitability of \PDSMNC in pretraining regression models for the dataset-adaptive workflow.

\label{sec:maxacc}

\subsubsection{Objectives and Study Design}

We want to evaluate the utility of \PDSMNC in training regression models that predict the maximum achievable accuracy of DR techniques. 
We assess the performance of regression models that predict the maximum accuracy achievable by DR techniques from the metric scores, comparing them to the ones that utilize baselines (intrinsic dimensionality metrics).
For 96 datasets that we have (\autoref{sec:acceval}), we first obtain maximum scores of DR techniques by optimizing them using Bayesian optimization (detailed settings in Appendix B). 
We then compute the $R^2$ scores of regression models predicting the maximum accuracy from \MNCPDS and baselines.
This is done by (1) splitting 96 datasets into 80 training datasets and 16 test datasets, (2) training the regression model with a training dataset with five-fold cross-validation, and (3) assessing the performance of the model for unseen test datasets.
To make scores comparable with previous experiments (\autoref{sec:acceval}), we repeat the measurement 10 times with different splits and report the ideal performance. 
For the structural complexity metric, we use \MNCPDS as it shows the best correlation with approximated ground truth structural complexity (\autoref{sec:acceval}).

\paragraph{Regression models}
To examine ideal prediction, we use AutoML~\cite{he21kbs} based on \texttt{auto-sklearn}~\cite{feurer15nips} to train the regression model. 
We train the model for 60 seconds. 
We do not test the five regression models used in our correlation analysis as they hardly achieve ideal predictions.

\paragraph{DR techniques and metrics}
We use the same set of DR techniques and evaluation metrics with the correlation analysis (\autoref{sec:acceval}).

\subsubsection{Results and discussions}
\autoref{tab:indmetric} depicts the results.
Overall, \MNCPDS has strong predictability with maximum accuracy achievable by DR techniques, outperforming baseline metrics.
For example, while \MNCPDS shows at least a strong predictive power for S-$\rho$ and P-$r$, the baselines fail to do so. 
\MNCPDS also has at least strong predictability for most combinations of DR techniques and evaluation metrics (26 out of 30; 87\%), verifying \revise{that the metric can be generally applied to execute the dataset-adaptive workflow in practice.} In contrast, projection-based and geometric projection metrics show at least strong predictability for 21 and 17 combinations, respectively. 

However, \MNCPDS relatively works poorly for L-T\&C, showing moderate or weak correlations for four cases. 
Such results align with the results from our correlation analysis (\autoref{sec:acceval}), clarifying the need for further development of complexity metrics that complement \PDS and \MNC. \PDSMNC still outperforms baselines also for L-T\&C.

\subsection{Evaluation on Predicting Effective DR Techniques}

\label{sec:evalguiding}

We investigate the effectiveness of \PDSMNC in predicting the accuracy ranking of DR techniques (Step 1; \autoref{fig:workflow} DW1).

\subsubsection{Objectives and Study Design}

\begin{figure}
    \centering
    \includegraphics[alt={A vertical bar chart showing rank correlations between various evaluation metrics and DR quality. The x-axis lists metrics: T&C, MRRE, L-T&C, S-ρ, and P-r. The y-axis ranges from -1.0 to 1.0. The chart includes interpretive labels: very weak, weak, moderate, strong, and very strong. It shows that PDS+MNC, Projection, and Geometric methods have significantly positive correlations (p < .05), especially for S-ρ and P-r.},width=\linewidth]{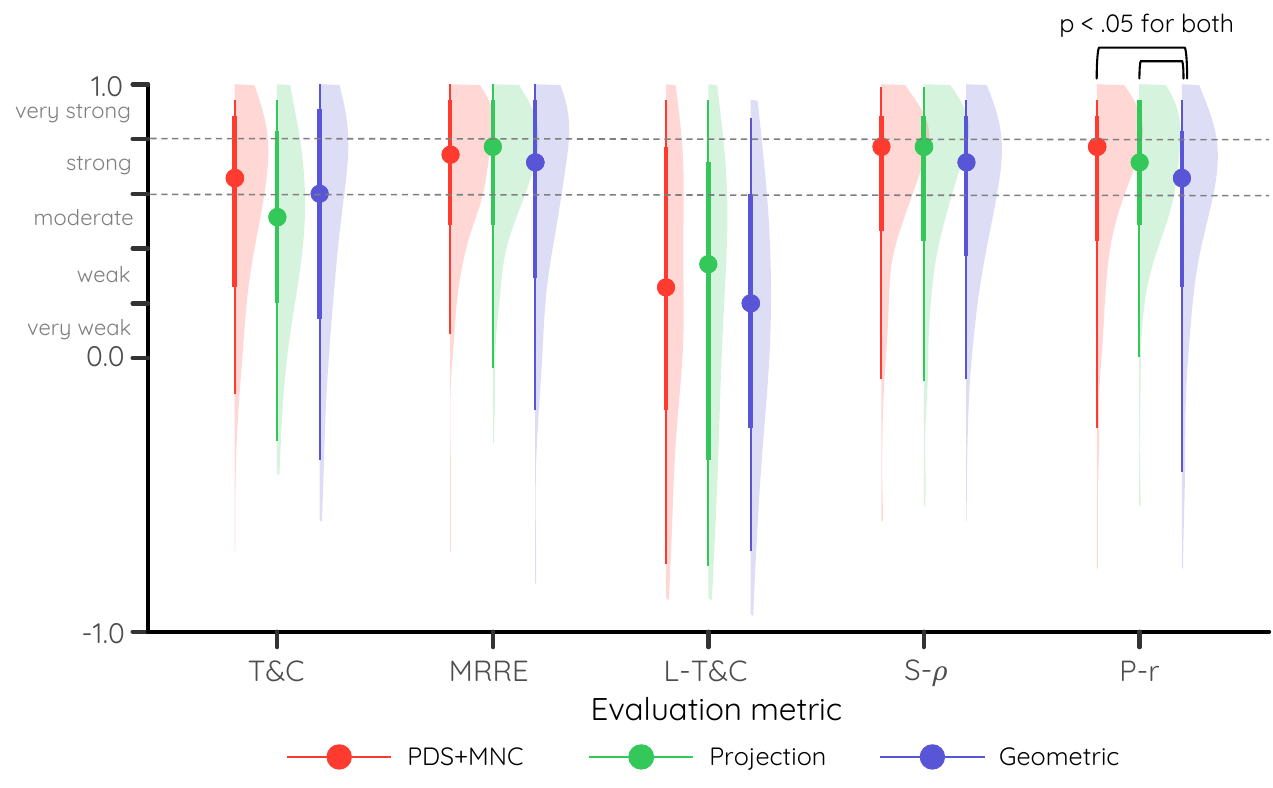}
    \caption{The distribution of the correlation between the true accuracy ranking of DR techniques and the predicted ranking estimated based on \MNCPDS and the baselines (\autoref{sec:evalguiding}). 
    We use the criteria of Prion and Hearing \cite{prion14csn} to explain the strength of the correlation (very strong to very weak).
    Overall, \PDSMNC demonstrates strong predictive power for the majority of cases.
    We also find that all three metrics have similar performance in predicting the ranking. The thick and thin lines indicate the 66\% and 99\% interval of probability mass.
    }
    \label{fig:correlation}
\end{figure}

We examine whether the pretrained regression models using \PDSMNC can effectively distinguish DR techniques with high and low maximum achievable accuracy in the testing phase.
This capability is important in properly guiding dataset-adaptive workflow to avoid the optimization of suboptimal DR techniques (\autoref{sec:datasetadaptive}, \autoref{fig:workflow} DW1).

We reuse the maximum achievable accuracy of DR techniques computed previously (\autoref{sec:maxacc}). Leveraging this score, we identify the ground truth accuracy ranking of DR techniques. We then train the regression models with 80 training datasets as we do in the previous experiment (\autoref{sec:maxacc}) using \PDSMNC and baselines (intrinsic dimensionality metrics). Based on the models’ approximated accuracy predictions, we derive the predicted accuracy ranking of DR techniques and compute the rank correlation between the ground truth and predicted rankings.
We run the same experiment 10 times with different splits of datasets and report the average results. 

\paragraph{DR techniques and metrics}
We use the same set of DR techniques and evaluation metrics with the previous experiments (\autoref{sec:acceval}, \ref{sec:maxacc}).

\subsubsection{Results and discussions}
We find that \PDSMNC achieve strong accuracy in predicting the ground truth ranking of DR techniques computed by T\&C, MRRE, S-$\rho$, and P-$r$ (\autoref{fig:correlation} red marks). The result reaffirms the capability of structural complexity metrics in guiding the dataset-adaptive workflow. In contrast, the rankings predicted by \PDSMNC show weak correlation with the ground truth for L-T\&C case, aligning with our previous experiments (\autoref{sec:acceval}, \ref{sec:maxacc}). 

We also find that baselines show good performance in predicting rankings that correlate well with ground truth rankings. We examine the difference in correlations between \PDS and two baselines for each evaluation metric using ANOVA. 
As a result, we find no significant difference for T\&C ($F_{2,477} = 2.26$, $p = .104$), MRRE ($F_{2,477} = 2.71$, $p = .067$), L-T\&C ($F_{2,477} = 1.48$, $p = .226$), and S-$\rho$ ($F_{2,477} = 0.91$, $p = .403$). 
We find a significant difference for P-$r$ case ($F_{2,477} = 5.37$, $p < .01$), and post-hoc test using Tukey's HSD confirms that \PDSMNC and projection-based intrinsic dimensionality metric significantly outperform geometric intrinsic dimensionality metric ($p < .05$ for both). Still, the geometric intrinsic dimensionality metric shows strong predictive power. 
These results indicate that distinguishing between effective and ineffective techniques does not require precise maximum accuracy predictions. We discuss the takeaways of this phenomenon \revise{on the practical use of dataset-adaptive workflow} in \autoref{sec:necessity}.

\subsection{Evaluation on Early Terminating Optimization}

\label{sec:evalguiding22}

We assess the utility of \PDSMNC and regression models in accelerating hyperparameter optimization by early terminating iterations.

\renewcommand{\hbar}[3]{%
  \begin{tikzpicture}[baseline=(textnode.base)]
    \pgfmathsetmacro{\barwidth}{0.87} 
    \pgfmathsetmacro{\barheight}{0.25}
    \pgfmathsetmacro{\perc}{#1/#2}
    \draw[blue!07, fill=blue!07] (0,0) rectangle (\barwidth,\barheight);
    \fill[blue!28] (0,0) rectangle (\perc*\barwidth,\barheight);
    \node (textnode) [anchor=mid, inner sep=0, font=\scriptsize] at (\barwidth/2, \barheight/2) {#3};
  \end{tikzpicture}%
}

\begin{table}[t]
    \centering
    \caption{The effectiveness of structural complexity metrics (\PDSMNC) and baselines (projection-based on geometric intrinsic dimensionality metrics) in guiding the early termination of hyperparameter optimization (\autoref{sec:enhanceeff}). We report the average error across all trials, the worst-case errors at the 10\%, 5\%, and 1\% percentiles, and the relative time required compared to full optimization. \revise{Each blue bar within a table cell represents the cell's relative value compared to the maximum value for each combination of DR evaluation metrics (rows) and measurands (columns).} The bold denotes the best performance for each combination.
    The results verify that the approximation using \PDSMNC consistently achieves desirable accuracy while maintaining efficiency. 
    }
    \scalebox{0.89}{
    \begin{tabular}{clccccc}
         \toprule 
                & & \multicolumn{4}{c}{Error} & \multirow{2.5}{*}{\makecell{Rel. \\Time}}\\
                 \cmidrule{3-6}
                &  & All & 10\% & 5\% & 1\% & \\
                \midrule
        \multirow{3}{*}{\makecell{UMAP \\(w\ T\&C)}} 
        & Projection   & \hbar{.0005}{.0005}{.0005} & \hbar{.0112}{.0157}{\textbf{.0112}} & \hbar{.0179}{.0192}{.0179} & \hbar{.0261}{.0285}{.0261} & \hbar{40.6}{40.6}{40.6\%} \\
        & Geometric    & \hbar{.0005}{.0005}{.0005} & \hbar{.0157}{.0157}{.0157} & \hbar{.0192}{.0192}{.0192} & \hbar{.0285}{.0285}{.0285} & \hbar{29.0}{40.6}{\textbf{29.0\%}} \\
        & \PDSMNC      & \hbar{.0004}{.0005}{\textbf{.0004}} & \hbar{.0118}{.0157}{.0118} & \hbar{.0159}{.0192}{\textbf{.0159}} & \hbar{.0246}{.0285}{\textbf{.0246}} & \hbar{35.7}{40.6}{35.7\%} \\
        \midrule
        \multirow{3}{*}{\makecell{UMAP \\(w\ P-$r$)}} 
        & Projection   & \hbar{.0395}{.0395}{.0395} & \hbar{.1287}{.1473}{.1287} & \hbar{.1560}{.1758}{.1560} & \hbar{.1913}{.2052}{.1913} & \hbar{50.3}{51.0}{50.3\%} \\
        & Geometric    & \hbar{.0380}{.0395}{.0380} & \hbar{.1473}{.1473}{.1473} & \hbar{.1758}{.1758}{.1758} & \hbar{.2052}{.2052}{.2052} & \hbar{44.7}{51.0}{\textbf{44.7\%}} \\
        & \PDSMNC      & \hbar{.0173}{.0395}{\textbf{.0173}} & \hbar{.0639}{.1473}{\textbf{.0639}} & \hbar{.0861}{.1758}{\textbf{.0861}} & \hbar{.1005}{.2052}{\textbf{.1005}} & \hbar{51.0}{51.0}{51.0\%} \\
        \bottomrule
    \end{tabular}
    }
    \label{tab:efferror}
\end{table}

\label{sec:enhanceeff}

\subsubsection{Objectives and Study Design}
We investigate the effectiveness of \PDSMNC in improving the efficiency of DR hyperparameter optimization by reducing redundant iterations (\autoref{fig:workflow} DW2), comparing with baseline metrics (intrinsic dimensionality metrics). 
We simulate the optimization of hyperparameters for UMAP projections. We evaluate using T\&C and P-$r$ to compare two scenarios: one in which \PDSMNC and baseline metrics exhibit similar performance (T\&C) and one in which \PDSMNC substantially outperforms the baselines in predicting the maximum achievable accuracy of DR techniques (P-$r$). 
This selection is based on our evaluation of pretraining regression models (\autoref{sec:maxacc}).
As with previous experiments, we divide the 96 HD datasets into 80 training datasets and 16 test datasets and train the AutoML regression model to predict the maximum accuracy from the training dataset. We then predict the maximum accuracy of 16 unseen test datasets. 
Note that for T\&C, we use the F1 score of Trustworthiness and Continuity, following the convention of interpreting T\&C as precision and recall of DR \cite{venna10jmlr}.

Finally, we optimize UMAP on 16 datasets with and without interruption based on the predicted maximum accuracy. 
For the former (with), we run Bayesian optimization with 50 iterations and halt the process upon reaching the optimal score. We set the default iteration number as 50, a default value recommended by \texttt{scikit-optimize} \cite{louppe2017bayesian} library. For the latter condition (without), we run the optimization with 50 iterations without halting the process. 
We compare the two settings by evaluating how much error is introduced and assessing the relative running time compared to the optimizations executed without early termination.
We run the same experiment 10 times with different splits of datasets.
Note that we report the average error over all trials and specifically examine errors from the worst 10\%, 5\%, and 1\% trials to detail the robustness of the early termination.

\subsubsection{Results and discussions}
\begin{figure}
    \centering
    \includegraphics[alt={A set of projection comparison results for three datasets: Fashion-MNIST, Optical Recognition, and Yeast/E. coli. Each dataset is visualized with and without early interruption. Text annotations include classification accuracy (Acc) and total time in seconds. For instance, Fashion-MNIST shows accuracy 0.979 (836.6s) without interruption vs. 0.983 (759.4s) with interruption. The trend suggests that early interruption reduces computation time while maintaining or improving accuracy.}, width=\linewidth]{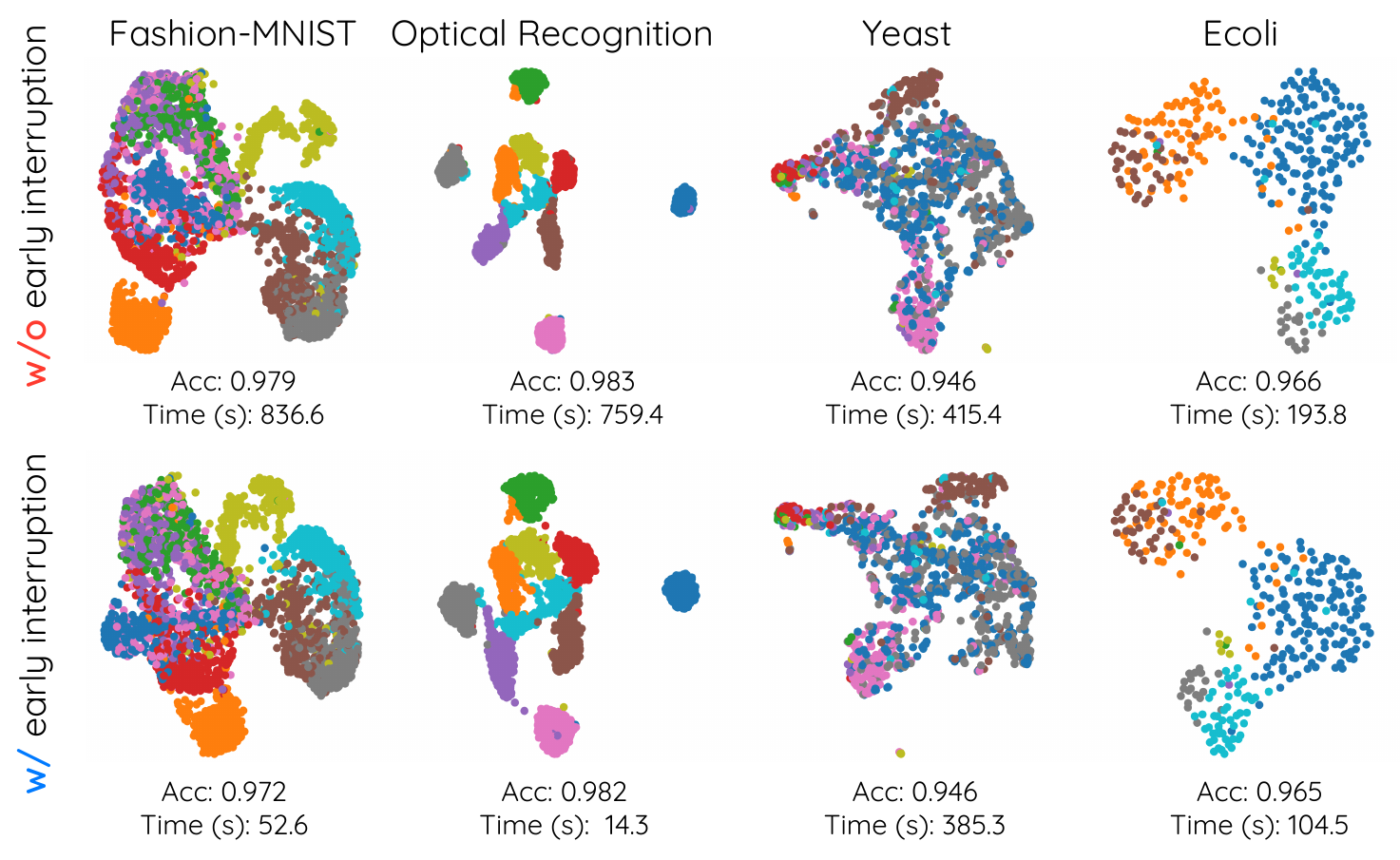}
    \caption{The projections made by optimizing UMAP with and without interruption based on predicted maximum accuracy (\autoref{sec:maxacc}).
    The interruption substantially reduce runtime while maintaining accuracy.}
    \label{fig:projections}
\end{figure}

\autoref{tab:efferror} depicts the results, and \autoref{fig:projections} depicts the subset of projections generated in our experiment.
The results demonstrate the efficacy of \PDSMNC in guiding the early termination of hyperparameter optimization, \revise{verifying its usefulness in executing the dataset-adaptive workflow in practice}. Specifically, early termination using \PDSMNC substantially reduces runtime while incurring only minimal errors. While reducing runtime by more than half, \PDSMNC outperforms the baselines in terms of error in the majority of cases. 
In the P-$r$ scenario, the error is approximately halved compared to the baselines.

\section{Experiment 3: \revise{Effectiveness of the Dataset-Adaptive Workflow}}

\label{sec:improvement}


We evaluate how well our dataset-adaptive workflow  (\autoref{sec:datasetadaptive}) accelerates the process of finding optimal DR projection without compromising accuracy, comparing it with conventional workflow (\autoref{sec:conventional}).


\subsection{Objectives and Study Design}
We aim to verify two hypotheses: 

\begin{itemize}
    \item[\textbf{H1}] The dataset-adaptive workflow significantly accelerates the DR optimization process compared to the conventional workflow.
    \item[\textbf{H2}] The dataset-adaptive workflow finds DR projections with negligible accuracy loss compared to the conventional workflow.
\end{itemize}
We first split 96 HD datasets into 80 training datasets and 16 test datasets. 
Then, we train regression models for all DR techniques we use previously (\autoref{sec:maxacc}). 
We then execute conventional (\autoref{sec:conventional}) and dataset-adaptive (\autoref{sec:datasetadaptive}) workflows for optimizing DR projection for test datasets, where we set the default iteration number as 50. 

For our dataset-adaptive workflow, we test two variants: one optimizes the hyperparameters of the top-1 DR technique, and the other optimizes those of the top-3 techniques. This is because we want to examine the tradeoff between the execution time and the accuracy. 
We record the total execution time and the final accuracy obtained by each workflow (conventional, top-1 dataset-adaptive, top-3 dataset-adaptive). We run the same experiment 10 times with diverse dataset splits.

\paragraph{Evaluation metrics}
We aim to examine both the full potential of the dataset-adaptive workflow and its effectiveness under worst-case scenarios. We measure the accuracy of DR projections using T\&C and L-T\&C, two metrics that \PDSMNC exhibits the best and worst predictive power in our previous experiment (\autoref{sec:maxacc}), respectively.


\subsubsection{Results and Discussions}

\begin{figure}
    \centering
    \includegraphics[alt={Two sets of dual-axis line plots comparing DR evaluation methods using Trustworthiness & Continuity (T&C) and Label-T&C over accuracy and execution time. Each subplot compares: Top-1 dataset-adaptive, Top-3 dataset-adaptive, and Conventional workflows. The x-axes represent accuracy (0.5 to 1.0), and y-axes show time (0 to 4000 sec). Statistical significance (p < .001) is noted for each comparison, emphasizing the efficiency and accuracy of dataset-adaptive approaches.}, width=0.5\textwidth]{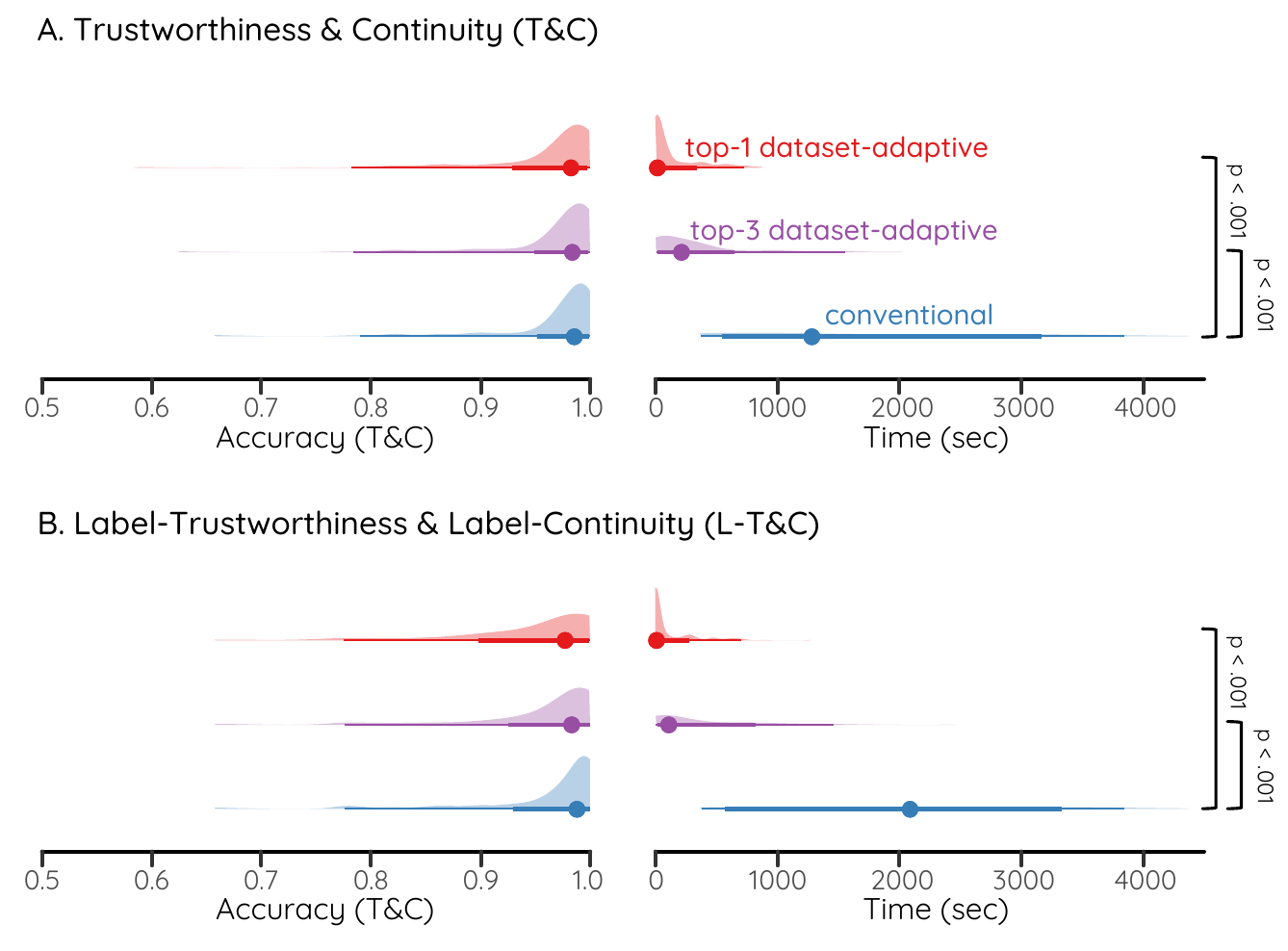}
    \caption{Comparison of the performance of three different workflows in optimizing DR projections. Dataset-adaptive workflows achieve a significant gain in execution time with negligible accuracy loss. The thick and thin lines indicate the 66\% and 99\% interval of probability mass.}
    \label{fig:fullpipe}
\end{figure}


\autoref{fig:fullpipe} depicts the results.
By running one-way ANOVA on the accuracy scores, we find no significant difference between workflows for both T\&C ($F_{2, 477} = 1.35$, $p = .259$) and L-T\&C ($F_{2, 477} = 2.49$, $p = .084$) cases. The result confirms \textbf{H2}.
In contrast, ANOVA on the execution time indicates the statistically significant difference between the workflows (T\&C: $F_{2, 477} = 194.18$, $p < .001$; L-T\&C: $F_{2, 477} = 190.97$, $p < .001$). 
We find that top-1 and top-3 dataset-adaptive workflows are $\times 5.3$ and $\times 13$ faster than the conventional workflow, respectively.
Post-hoc analysis using Tukey's HSD also confirms that top-1 and top-3 dataset-adaptive workflows require significantly shorter execution time than the conventional workflow ($p < .001$). 
This result confirms \textbf{H1}. 
Meanwhile, we observe no significant difference between the top-1 and top-3 dataset-adaptive workflows (T\&C: $p = .056$, L-T\&C: $p = .077$).
In summary, our experiment verifies both hypotheses, confirming the benefit of using the dataset-adaptive workflow in practice. 

The results also reveal that the dataset-adaptive workflow does not always perform ideally. Although the accuracy differences are not statistically significant, the L-T\&C case yields a relatively low $p$-value (0.084), indicating a slight accuracy difference in the conventional and the dataset-adaptive workflow.
Such results suggest that more efforts should be invested in designing advanced structural complexity metrics and further refining the dataset-adaptive workflow.

\section{Discussions}

We discuss future research directions for dataset-adaptive workflow and structural complexity metrics.

\subsection{Leveraging the Tradeoffs between Predictive Power and Efficiency}

\label{sec:necessity}



We find that \PDSMNC outperforms intrinsic dimensionality metrics in early terminating hyperparameter optimization (\autoref{sec:evalguiding22}) but is similarly effective in identifying optimal DR techniques (\autoref{sec:evalguiding}). This indicates that the predictive power of structural complexity metrics is crucial for the success of the former but not for the latter.

This finding offers a new perspective on designing structural complexity metrics: leveraging the tradeoff between predictive power and efficiency. Although \PDS and \MNC run within a reasonable time frame for our dataset, they may be too slow for enormous datasets with millions of data points. 
In such cases, we can use fast but less accurate structural complexity metrics to select the DR technique in Step 1 and then apply accurate metrics like \PDS and \MNC only to the selected technique in Step 2.
\revise{We can also dynamically adjust the optimization procedure for multiple DR techniques. For example, we can run multiple DR techniques in parallel and drop those predicted to produce low values. Applying a progressive visual analytics paradigm \cite{fekete24book} will be effective here. For instance, we can begin optimization with rough estimates of promising DR techniques and eliminate those that prove to be ineffective later in the process.}

\revise{To pursue this direction, it will also be crucial to investigate the utility of structural complexity metrics in depth. For example, we may examine the gap between the achieved predictive power and the theoretical optimum, and whether this gap is bounded. When metrics are computed progressively, it will also be important to investigate how the error bounds evolve over time.}
These endeavors will contribute to making DR-based visual analytics both more responsive and accurate. 


\subsection{Complementing Our Structural Complexity Metrics}

Our experiments reveal the necessity of developing new structural complexity metrics that complement \PDS and \MNC, especially the ones that focus on cluster-level structure. This is because \PDS, \MNC, and even \PDSMNC fall short in estimating the ground truth structural complexity and the maximum accuracy of DR techniques computed with L-T\&C, a cluster-level evaluation metric. One idea for developing cluster-level structural complexity metrics is to extend \MNC to the class level. Instead of considering neighbors of ``data points,'' we could consider those of ``classes'' or ``clusters.''




\subsection{Exploring Additional Use Cases}

We hypothesize that structural complexity metrics can improve the \textit{replicability} of benchmarking DR techniques.
DR benchmarks may produce different conclusions about the accuracy of DR techniques depending on the datasets used, i.e., they may have low replicability.
Adding more benchmark datasets may enhance replicability, but this increases the computational burden. 
Here, complexity metrics may contribute to achieving high replicability with fewer datasets.
For example, excluding datasets with simple patterns that are accurately reducible by any DR technique will improve replicability, as the rankings of DR techniques determined based on these datasets may noise the evaluation.
Validating whether this hypothesis confirms or not will be a worthwhile future research avenue to explore.

\section{Conclusion}

It is important to find optimal DR projections to ensure reliable visual analytics. 
However, optimization processes are computationally demanding. 
We propose the dataset-adaptive workflow for accelerating DR optimization while maintaining accuracy. 
We introduce two structural complexity metrics, \PDS and \MNC, and verify their effectiveness in terms of precision and efficiency. 
We also demonstrate the utility of the dataset-adaptive workflow guided by these two metrics.
Overall, our proposal opens up the discussion towards achieving a more reliable and efficient analysis of HD data.

\section*{\revise{Online Supplemental Materials}}

\revise{
We release our supplemental materials online (\href{https://hyeonword.com/dadr}{\texttt{hyeonword.com/dadr}}). This includes the Appendix and the code repository to reproduce our dataset-adaptive workflow.
}

\acknowledgments{%
This work was supported by the National Research Foundation of
Korea (NRF) grant funded by the Korean government (MSIT) (No.
2023R1A2C200520911), the Institute of Information \& communications Technology Planning \& Evaluation (IITP) grant funded by the Korean government (MSIT) [NO.RS-2021-II211343, Artificial Intelligence Graduate School Program (Seoul National University)], and by the SNU-Global Excellence Research Center establishment project. The ICT at
Seoul National University provided research facilities for this study.
Hyeon Jeon is in part supported by Google Ph.D. Fellowship. 
The authors thank Hyunwook Lee for providing insightful feedback that helped improve this work.}

\bibliographystyle{abbrv-doi-hyperref}

\bibliography{ref}










\end{document}